\documentclass[a4paper]{article}
\title{
Anderson Theorem for Color Superconductor\\}
\author{B.O. Kerbikov\\ State Research
Center\\Institute of Theoretical and Experimental Physics, \\
Moscow, Russia}
 \date{}
  \newcommand{\be}{\begin{equation}}
\newcommand{\ee}{\end{equation}}
 
\def\fun#1#2{\lower3.6pt\vbox{\baselineskip0pt\lineskip.9pt
\ialign{$\mathsurround=0pt#1\hfil ##\hfil$\crcr#2\crcr\sim\crcr}}}

\newcommand{\bpsi}{\bar\psi}

\newcommand{\vep}{\mbox{\boldmath${\rm p}$}}

\newcommand{\vepi}{\mbox{\boldmath${\rm \pi}$}}

 \newcommand{\llan}{\langle\langle}
\newcommand{\rran}{\rangle\rangle}

\begin{document}

\maketitle

\begin{abstract}

Ginzburg-Landau functional is derived for a system possessing both
chiral and diquark condensates. Anderson theorem for such a system
is formulated and proved.
\end{abstract}

\section{Introduction}
During the last couple of years color superconductivity became a
compelling topic in QCD - see e.g. the review papers \cite{1,2}.
The present work is aimed to shed some new light on the region of
mixed condensation in which both chiral and color symmetries are
broken. For definiteness we consider the case of two flavors $u$
and $d$, the corresponding color superconducting phase is called
2SC \cite{1,2}. The point at issue is the interplay of the chiral
condensate $\varphi$ and the diquark condensate $\Delta$ in the
interval of the chemical potential $\mu$ where they possibly
coexist. Necessary to mention that as model calculations show
\cite{3}-\cite{6} the  very presence of such a region is
questionable. Instead of coexistence one  rather observes a strong
competition between $\varphi$ and $\Delta$ in the sense that where
one condensate is nonzero the other vanishes. This conclusion has
been drawn as a result of numerical  analysis of the   complicated
set of two coupled gap equations for $\varphi$ and $\Delta$
\cite{3,5} (other authors take it  as a plausible apriori
assumption \cite{4}).
 The question has been also addressed within the random matrix approach \cite{7a}
 with the  result  that for single-gluon exchange
   the coexistence region of the two condensates is absent.
  Evidently such a situation calls for
simplification. Arguments presented below are aimed to show that
as soon as (with $\mu$ increasing) the diquark condensate $\Delta$
develops the role of the chiral condensate $\varphi$ is prescribed
by what may be called the Anderson theorem \cite{7,8}. In BCS
theory this theorem states that nonmagnetic impurity do not
influence the thermodynamic properties of a superconductor (in
particular the values  of the gap and of the critical temperature)
in linear approximation in impurity concentration. In case of
color superconductor the role of impurity is taken by chiral
condensate. This means that  the system of the gap equations for
$\varphi$ and $\Delta$ actually decouples and with high accuracy
one can calculate $\Delta$ neglecting the influence of $\varphi$.
One general remark is probably needed. Namely, the  content of the
"true" Anderson theorem  in BCS is physically deeper than the
statements formulated below. This point has to be discussed within
comparative study of BCS theory and color superconductivity which
is beyond our present goals (see [1,2]).

  In Section 2 we
present and discuss the general expression for  thermodynamic
potential which has the structure common to three approaches to
color  superconductivity, namely NJL, one gluon exchange, and
instanton ensemble. In Section 3 we  derive the Ginzburg-Landau
functional for chiral condensate, while in Section 4 the same is
done  for that part of the potential which depends on both
condensates - the diquark and the chiral ones. The main result of
this Section is that the contribution of the chiral condensate to
the "mixed" part of thermodynamic potential may appear only in
orders higher than second one.  In Section 5 we show that  the
same is also true at zero temperature. Section 5 is devoted to
the formulation and discussion of what we call the Anderson
theorem for color superconductor.

\section{Partition Function and Thermodynamic Potential}

We start with the QCD Euclidean partition function
\be
Z=\int DAD\bpsi D\psi \exp(-S),\label{1} \ee where
\be
S=\frac{1}{4} \int F^a_{\mu\nu} F^a_{\mu\nu} d^4 x=\int\bpsi
(-i\gamma_\mu D_\mu-im +i\mu\gamma_4)\psi d^4 x.\label{2}\ee

In (\ref{2}) color and flavour indices are suppressed,
$N_f=2,~N_c=3$, and the chemical potential $\mu$ is introduced.
Performing integration over the gauge fields one gets effective
fermion action in terms of cluster expansion
\be
Z=\int D\bpsi D\psi \exp(-\int d^4xL_0-S_{eff}),\label{3}\ee with
$L_0=\bpsi(-i\gamma_\mu\partial_\mu-im+i\mu\gamma_4)\psi$ and
effective action $S_{eff}=\sum^\infty_{n=2}
\frac{1}{n!}\llan\theta^n\rran$, where $\theta=\int d^4x\bpsi(x)
g\gamma_\mu A^a_\mu(x)t^a\psi (x)$ and double brackets denote
irreducible  cumulants \cite{9}. Since we consider only $u$ and
$d$ quarks    the current quark mass $m$ will be neglected. In
future  we plan to give up this approximation especially in view
of the numerical results obtained in \cite{3} which show that
there is some influence of $m$ on the phase structure near the
transition point. One should also keep in mind that integration
over gauge fields when passing from (\ref{1}) to (\ref{3}) is in
no way a  trivial operation. It certainly deserves dedicated
investigation which is outside the scope of the present work (for
some guidlines see \cite{10}).

To proceed further some simplification of (\ref{3}) is needed.
Being quite general, (\ref{3})  gives rise to several commonly
used models. First step is to keep in $S_{eff}$ only the lowest
four-quark interaction. Routine manipulations then lead to the one
gluon exchange model \cite{11} which in turn displays the bulk
properties of the color superconductivity phenomena \cite{1}. One
gluon exchange yields surprisingly interesting results at very
high density \cite{1,2}. Less trivial but very transparent
approach \cite{12} enables to recast (\ref{3}) into the delute
instanton gas model which is successfully used in color
surerconductivity studies \cite{13,5,1}. We note  that derivation
of the instanton model presented in \cite{12} goes beyond zero
modes approximation. The importance of higher modes was
demonstrated in \cite{14,15}. Simplifying things even further
\cite{12} one arrives at the NJL models which as also quite
suitable to describe color superconductivity \cite{1,4}.

A lesson from comprehensive studies of color superconductivity
within different models listed above is that the main results are
very similar \cite{1,2}. The key technical points leading from
chosen Lagrangian to the effective action are similar as well. Use
is made of the bosonization of the fields $(\bar \psi \psi) $ and
($\psi\psi)$ \cite{3}. As a result the following general
expression in the Nambu-Gorkov basis of eight component fields
$(\psi,\bpsi^T)$ emerges:
\be
 S_{eff}=\int d^4x\left\{
\frac{\varphi^2}{4g^2_1}+\frac{\Delta^+\Delta}{4g^2_2}-\frac12
tr\ln\left(\begin{array}{ll} \Delta \Phi&
i\partial_\mu\gamma_\mu+i\varphi\Lambda-i\mu\gamma_4\\
-i\partial^T_\mu\gamma^T_\mu-i\varphi\Lambda^T+i\mu\gamma_4&\Delta^+\Phi^+\end{array}
\right)\right\}.\label{4}\ee Let us present necessary explanations
to (\ref{4}). We use the following representation
\be
\gamma_k=\left(\begin{array}{ll} 0&-i\sigma_k\\
i\sigma_k&0\end{array}\right),~~ \gamma_4=\left(\begin{array}{ll}
1&0\\ 0&-1\end{array}\right)\label{5}\ee
\be
C=\gamma_2\gamma_4,~~ C^{-1}\gamma_\mu
C=-\gamma_\mu^T,\partial^T_\mu=
\overleftarrow{\partial}_\mu=-\overrightarrow{\partial}_\mu.\label{6}\ee
Effective action (\ref{4}) is written in terms of the two
condensates - the chiral one $\varphi$ and the superconducting
diquark one $\Delta$. Operators $\Phi$ and $\Lambda$ in (\ref{4})
determine the matrix structure of these condensates
\be
\Delta^{\alpha\beta}_{ij}=\varepsilon_{\alpha\beta
3}\varepsilon_{ij} C\gamma_5 \Delta\equiv \Phi\Delta,~
\varphi^{\alpha\beta}_{ij}=\delta_{\alpha\beta}\delta_{ij}\varepsilon
\equiv \Lambda\varphi,\label{7}\ee where indices $\alpha,\beta$
correspond to color, and $i,j$ - to flavor.
 The matrix structure
defined by (\ref{7}) is the simplest possible. For example, if one
starts with the single gluon exchange model and performs Fierz
transformation with respect to Lorentz, color and flavor indices,
one arrives to several other diquark condensates in addition to
(\ref{7}), e.g. the negative parity one with
$\Phi'=\varepsilon_{\alpha\beta 3} \varepsilon_{ij}C$, etc., see
\cite{1,16}. The same is true for the matrix $\Lambda$. The
minimal "improvement" of $\Lambda$ in line with chiral invariance
requirement would be $\Lambda'=\tau_{a=0-3}+i\gamma_5\tau_{a=0-
3}$ while in (\ref{7}) only the $a=0$ part of the first term in
kept. In other words from the two fields $(\sigma,\vepi)$ only
$\sigma$ is  kept.
 The two
coupling constants $g^2_1$ and $g^2_2$ have the dimension of
$m^{-2}.$ In the one gluon exchange inspired model they are
related to each other  but in most studies they are considered as
independent \cite{1}.  Regularization of the integrals involving
the chiral gap is performed either by smooth formfactor or by
cutoff. The sensitivity to the regularization procedure is
unessential \cite{1} and to avoid  complicated equations we do not
introduce this  regularization  explicitly. Integrals involving
the superconducting  gap $\Delta$ are regularized at the Debye
frequency --see below.

Factorizing from (\ref{4}) the four-volume and making use of  the
identity
\be
tr\ln\left(
\begin{array}{ll}
R&A\\ -A^T& R^+\end{array}\right)= tr\ln AA^T+ tr\ln\left\{
1+A^{-1} R(A^T)^{-1}R^+\right\}\label{8}\ee one arrives at the
following expression for thermodynamic potential
$\Omega(\varphi,\Delta; \mu, T)$ $$ \Omega(\varphi,\Delta:\mu,
T)=\frac{\varphi^2}{4g^2_1} +\frac{\Delta^+\Delta}{4g^2_2}- tr \ln
(-\hat p +i\varphi\Lambda-i\mu\gamma_4)-$$
\be
-\frac{1}{2} tr\ln\left\{ 1+\frac{\Delta\Delta^+\Phi\Phi^+}{(\hat
p_+-i\varphi\Lambda)(\hat p_-+i\varphi
\Lambda)}\right\},\label{9}\ee where $\hat p= p_\mu\gamma_\mu,
p_{\pm}=(p_k,p_4\pm i\mu)$. This form on $\Omega $ was first
derived in \cite{3}.

The remaining part of the paper will be mostly devoted to the
investigation of $\Omega$ in the Ginzburg-Landau region, i.e. to
the expansion of $\Omega$ in powers of $\Delta$ and $\varphi$ at
temperatures close to the critical one. Obviously the system under
consideration  possesses two different critical temperatures - the
chiral temperature $T_\varphi$ and the diquark one $ T_\Delta$.
The situation simplifies if one focuses on the transition region
in the $(T,\mu)$ plane where the two condensates $\varphi$ and
$\Delta$ compete. Numerical calculations show \cite{3} that in the
corresponding interval of $\mu$ the slope $\frac{dT}{d\mu}$ of the
phase transition line is small and  a notion of a common critical
temperature $T_c$ makes sense. At $T>T_c$ both condensates
$\varphi$ and $\Delta$ vanish. Numerically $T_c$ is about 30-40
MeV while the corresponding values of $\mu$ lie in the interval of
300-400 MeV \cite{3}. This is the region of the phase diagram
which we will be interested in. Consideration will be restricted
to the case of homogeneous condensates $\varphi$ and $\Delta$.
Thus the gradient terms will be absent.

We start the derivation with the expansion of the third term in
(\ref{9}) in powers of $\varphi$.

\section{Chiral Condensate Transition  Line}

Consider the "$\varphi$-part" of the partition function (\ref{9}),
i.e. the functional
\be
\Omega_\varphi (\varphi;\mu, T)=\frac{\varphi^2}{4g^2_1}
-tr\ln(-\hat p+i\varphi\Lambda-i\mu\gamma_4).\label{10}\ee

The equation of $\Omega_\varphi$ proceeds along the standard lines
\cite{17}. One has
\be
\Omega_\varphi=\frac{\varphi^2}{4g^2_1}+2N_cN_fT\int
d\varphi^2\sum_n\int \frac{d^3p}{(2\pi)^3}\frac{1}{[(2n+1)\pi
T+i\mu]^2+ p^2+\varphi^2},\label{11}\ee where $N_c=3, N_f=2$, and
the summation is over the fermionic Matsubara modes. After
performing the sum and $d\varphi^2$ integration the result for
$\Omega_\varphi$ reads
\be
\Omega_\varphi =\frac{\varphi^2}{4g^2_1}-\frac{N_c N_f}{\pi^2}
T\int dp p^2(\ln 2ch\theta^{(-)}+\ln
2ch\theta^{(+)}),\label{12}\ee \be
\theta^{(\mp)}=\frac{\sqrt{p^2+\varphi^2}\mp\mu}{2T}.\label{13}\ee

Next we expand $\Omega_\varphi$ in powers of $\varphi$ in the
vicinity of $T_c$. Expansion of the thermodynamic potential
containing the second and the fourth order terms is called
Ginzburg-Landau functional. The first order term is absent because
of the mean-field stationarity condition
$\partial\Omega_\varphi/\partial\varphi=0$. The discussion of the
interesting physics associated with the tricritical point and
sixth order term is out of the  scope of the present paper, see
\cite{1,2,18,19}.

With the expression (\ref{12}) for the thermodynamic potential at
hands the calculation of the derivatives
$\partial^2\Omega_\varphi/\partial\varphi^2$ and
$\partial^4\Omega_\varphi/\partial\varphi^4$ is rather
straightforward. One has

\be
2 A(T)\equiv\left. \frac{\partial^2\Omega_\varphi}{\partial
\varphi^2} \right|_{\varphi=0} =\frac{1}{2g^2_1}-\frac{N_c
N_f}{2\pi^2} \int dpp(th\theta^{(-)}+th \theta^{(+)}).\label{14}
\ee

Denoting $\theta^{(\mp)}$ at $T=T_c$ as $\theta_c^{(\mp)}$ one
writes
\be
th\theta^{(\mp)}\simeq th \theta^{(\mp)}_c
+\left(\frac{T_c-T}{2T^2_c}\right)\frac{(p\mp\mu)}{ch^2\theta_c^{(\mp)}}.\label{15}\ee
Substitution of (\ref{15}) into (\ref{14}) leads after simple
integration to the result
\be
A(T)=\frac{1}{6} N_cN_fT^2_c\left(\frac{T-T_c}{T_c}\right)
=T^2_c\left(\frac{T-T_c}{T_c}\right).\label{16}\ee

In arriving to (\ref{16}) use was also made of the gap equation
$\partial\Omega_\varphi/\partial\varphi=0$ which yields
cancellation of the $1/2g^2_1$ term. Strictly speaking the gap
equation is $\partial\Omega/\partial\varphi=0$, where $\Omega$ is
the full thermodynamic potential (\ref{9}) but as we shall see the
corresponding corrections are small. Note that $A<0$ at $T<T_C$ in
line with the general theory of phase transitions.

Calculation  of the fourth order derivative is almost as simple.
As a preliminary remark we remind that for $N_f=2$ chiral symmetry
restoration at $\mu=0$ proceeds via second order phase transition.
On the other hand at $T=0$ but with $\mu$ increasing the
restoration of chiral symmetry occurs via first order phase
transition \cite{1,2,18,19}. Correspondingly in the  first case
the coefficient of the quartic term has  to be positive while in
the second case -negative. The sign is changed at the tricritical
point \cite{1,2,18, 19} at which the sixth order term is
important. Consideration of this point is beyond the scope of the
present paper. Another remark is that in the NJL model the order
of the phase transition depends upon the value of the cutoff
\cite{20}. We shall immediately see that indeed the sign of the
fourth order term depends on the value of the chemical-potential
$\mu$ and of the cutoff value $\Lambda$.

The nonvanishing at  $\varphi=0$ part of the fourth order
derivative reads

\be
4B(T,\mu)\equiv\frac{\partial^4\Omega_\varphi}{\partial\varphi^4}=\frac{3N_cN_f}{2\pi^2}\int
dp\left\{\frac{th
\theta^{(-)}+th\theta^{(+)}}{p}-\frac{1}{2T_c}\left(\frac{1}{ch^2\theta^{(-)}}
+\frac{1}{ch^2\theta^{(+)}}\right)\right\}.\label{17}\ee

Simple manipulations yield
\be
B(T,\mu)=\frac{9}{2\pi^2}\left\{ \int^{\Lambda/T_c}_0 dt \frac{th
t}{t}\left(
\frac{cht}{cht+ch\frac{\mu}{T_C}}\right)-1\right\}.\label{18}\ee
As expected  the sign of $B$ depends on the values of the chemical
potential $\mu$ and cutoff $\Lambda$. Having in mind to
demonstrate explicity the dependence on the cutoff we may resort
to a rough estimate, namely $$
 \int^{\Lambda/T_c}_0 dt \frac{th
t}{t}\left(
\frac{cht}{cht+ch\frac{\mu}{T_C}}\right)<\int^{\Lambda/T_c}_0 dt
\frac{th t}{t}= th\frac{\Lambda}{T_c}\ln \frac{\Lambda}{T_c}
-\int^\infty_0 dt\frac{\ln t}{ch^2t}= $$
\be
= th\frac{\Lambda}{T_c}\ln \frac{\Lambda}{T_c}+\ln
\frac{4\gamma}{\pi}\simeq\ln\frac{4\gamma\Lambda}{\pi
T_c},\label{19}\ee where $\gamma=e^C,C=0.5777$... Typical cutoff
value is $\Lambda\sim 800 $ MeV \cite{1,2,20}.

Collecting pieces together we write  down the Ginzburg-Landau
functional for the chiral part of the thermodynamic potential
$(N_c=3,N_f=2)$;
\be
\Omega_\varphi=T^2_c\left(\frac{T-T_c}{T_c}\right)\varphi^2+\frac{9}{2\pi^2}\left\{
 \int^{\Lambda/T_c}_0 dt \frac{th
t}{t}\left(
\frac{cht}{cht+ch\frac{\mu}{T_C}}\right)-1\right\}\varphi^4.\label{20}\ee

We note that the above treatment was oversimplified in several
points, in particular we assumed that the coupling $g^2_1$ was
temperature independent, the quark current masses were set to
zero, etc. These omissions important by themselves are hardly
crucial for our main purpose which is the formulation and proof of
the Anderson theorem.

\section{Ginzburg-Landau potential for diquarks}

Now we turn to the last term in Eq. (\ref{9}). Expanding the
logarithm in powers of $\Delta^2$ we directly arrive at the
desired Ginzburg-Landau functional. The term proportional to
$\Delta^2$ reads
\be
\Omega^{(2)}_{\Delta\varphi} =\frac{\Delta^2}{4g^2_2}-\frac12 tr
\frac{\Delta^2 \Phi\Phi^+}{(\hat p_+-i\varphi\Lambda)(\hat
p_-+i\varphi\Lambda)}.\label{21}\ee

Inverting the denominator and performing the trace over discrete
indices one gets $(N_c=3)$
\be
\Omega^{(2)}_{\Delta\varphi} =\frac{\Delta^2}{4g^2_2}- 8\Delta^2
T\sum_n\int\frac{d^3p}{(2\pi)^3}
\frac{(p^2_4+\vep^2+\mu^2+\varphi^2)}{R}, \label{22}\ee where
\be
R=p^4_4+2p^2_4(\vep^2+\mu^2+\varphi^2)+(\vep^2-\mu^2+\varphi^2)^2,\label{23}\ee
and the sum  (\ref{22})  is taken over fermionic Matsubara modes
corresponding to $p_4$. Next step differs from the calculation of
the $\varphi^2$ term presented in the previous section. Namely we
assume that considering the Landau-Ginzburg region for the diquark
condensate $\Delta$ one can neglect the contribution of
antiparticles. This means that in the representation
\be
\frac{p^2_4+\vep^2+\mu^2+\varphi^2}{R} =\frac12\left\{
\frac{1}{p^2_4+(\sqrt{\vep^2+\varphi^2}-\mu)^2}+
\frac{1}{p^2_4+(\sqrt{\vep^2+\varphi^2}+\mu)^2}\right\}\label{24}\ee
only the first term is kept. Then summation in (\ref{22}) yields
\be
\Omega^{(2)}_{\Delta\varphi}=\frac{\Delta^2}{4g^2_2}-\Delta^2N(0)\int^{\omega_D}_{-\omega_D}
d\xi\frac{1}{q}th\frac{q}{2T},\label{25}\ee where \be
\xi=p-\mu,~~q=\sqrt{p^2+\varphi^2}-\mu,\label{26}\ee and
$\omega_D$ is the Debye frequency while $N(0)$ is the density of
states at the Fermi surface
\be
N(0)=\frac{2}{\pi^2}\left(p^2\frac{dp}{d\xi}\right)_F=\frac{2\mu^2}{\pi^2}.\label{27}\ee
The density of states (\ref{27}) is four times larger than the BCS
theory factor $\mu^2/2\pi^2$. This is due to color and flavor
degrees of freedom, in 2SC there are red and green $u$ and $d$
quarks.

Next we expand the integrand in (\ref{25}) in powers of ($T-T_c)$
and $\varphi$. This  gives\be \frac{1}{q} th\frac{q}{2T}\simeq
\frac{1}{\xi} th
\frac{\xi}{2T_c}+\left(\frac{T_c-T}{4T^2_c}\right)
ch^{-2}\frac{\xi}{2T_c}+\frac{\varphi^2}{8\mu T^2_c}
\left(\frac{1}{y} th y\right)', \label{28}\ee with $y=\xi/2T$. The
last  term drops upon integration over the interval
$[-\omega_D/2T_c , \omega_D/2T_c]$. Integration results also in
cancellation of the first term with the term $\Delta^2/4g^2_2$ in
(\ref{21}) (because of the gap equation or equivalently from the
definition of the critical temperature). Thus
\be
\Omega^{(2)}_\Delta= \Delta^2
N(0)\left(\frac{T-T_c}{T_c}\right).\label{29}\ee We see that up to
$\varphi^2$ there is no dependence on the chiral condensate,
therefore we dropped the index $\varphi$ in (\ref{29}). At this
point one may notice the analogy with the Anderson theorem in BCS
theory.

Next we consider the term proportional to $\Delta^4$ in the
expansion of the logarithm in (\ref{9}). In this term the
$\varphi$ dependence will be neglected from the beginning. There
is no need to present the details of the calculations which are
simple and essentially  repeat the derivation of the $\Delta^2$
term with only two new technical points which are
\be
\sum_n\frac{1}{(\omega^2_n+\varepsilon^2)^2}=-\frac{1}{2\varepsilon}
\frac{\partial}{\partial\varepsilon}
\sum_n\frac{1}{\omega^2_n+\varepsilon^2},~~\int^\infty_0\frac{dy}{y}\left(\frac{thy}{y}\right)'
=-\frac{7\zeta (3)}{\pi^2}.\label{30}\ee

The result for  $\Omega^{(4)}_\Delta$ reads
\be
\Omega_\Delta^{(4)}= \Delta^4
N(0)\frac{7\zeta(3)}{16\pi^2T_c^2}.\label{31}\ee

Thus the development of $\Omega_{\Delta\varphi}$ in powers of
$\Delta$ and $\varphi$ reads
\be
\Omega_\Delta=-\Delta^2 N(0)\left\{
1-\frac{T}{T_c}-\frac{7\zeta(3)}{16\pi^2
T^2_c}\Delta^2\right\}.\label{32}\ee

We note  that with  realistic values of $\Delta$ and $\mu$
\cite{1}  the numerical value of $\Omega_{\Delta\varphi}$ given by
(\ref{32}) is in perfect agreement  with our estimate presented in
\cite{11}.

Now we are  tooled to formulate conclusions concerning the
interplay of the two condensates. However we first turn from the
Ginsburg-Landau region to the $T=0$ case in order to demonstrate
that the version of the Anderson theorem also holds at $T=0$.

\section{Renormalized Potential at $T=0$}

At $T=0$ the formal development of the thermodynamic potential
(\ref{9}) in powers of $\Delta$ and $\varphi$ is not appropriate.
Instead we calculate the potential explicitly. The
"$\varphi$-part" of it was evaluated in Sec. 3 exactly so we may
just set $T=0$ in (\ref{12}) and  get
\be
\Omega_\varphi(T=0)=\frac{\varphi^2}{4g^2_1}-\frac{N_cN_f}{\pi^2}\int
dpp^2\{\sqrt{p^2+\varphi^2}+
(\mu-\sqrt{p^2+\varphi^2})\theta(\mu-\sqrt{p^2+\varphi^2})\}.\label{33}\ee

Next we turn to the last term in (\ref{9})
\be
\Omega_{\Delta\varphi} =\frac{\Delta^2}{4g^2_2} -\frac12 tr\ln
\left\{ 1+\frac{\Delta \Delta^+\Phi\Phi^+}{(\hat p_+-i\varphi
\Lambda)(\hat p_-+i\varphi\Lambda)}\right\}.\label{34}\ee

Here use can be made of the standard trick; first one takes
$\Delta\Omega_{\Delta\varphi}/\partial\Delta$, then performs
integration over $dp_4$ which  corresponds to $T=0$, and then
reconstructs $\Omega_{\Delta\varphi}(T=0)$ by integration back
over $\Delta$. The result  reads
\be
\Omega_{\Delta\varphi}
(T=0)=\frac{\Delta^2}{4g^2_2}-\frac{2}{\pi^2}\int
dpp^2\{\sqrt{(\sqrt{p^2+\varphi^2}-\mu)^2+\Delta^2}+
\sqrt{(\sqrt{p^2+\varphi^2}-\mu)^2+\Delta^2}\}.\label{35}\ee

As before the second term corresponding to antiparticles will be
neglected. In order to avoid the onset of "irrelevant" terms of
higher order in $\Delta$ we renormalize $\Omega_{\Delta\varphi}
(T=0)$ at certain scale $\kappa\gg\mu$ \cite{21}. The renormalized
coupling constant $G^2_2$ is defined as
\be
\frac{1}{G^2_2}\equiv\left|\frac{\partial^2\Omega_{\Delta\varphi}(T=0)}{\partial\Delta^2}\right|
{\Delta=\kappa}.\label{36}\ee

The renormalized potential reads
\be
\Omega^r_{\Delta\varphi} =\frac{\Delta^2}{2G^2_2} -N(0)\int
dp\{R_\Delta-\frac{\Delta^2}{2R_\kappa}+\frac{\Delta^2\kappa^2}{2R^3_{\kappa}}\},\label{37}\ee
where
\be
R_\alpha=\{(\sqrt{p^2+\varphi^2}-\mu)^2+\alpha^2\}^{1/2},~~
\alpha=\Delta, \kappa.\label{38}\ee

Performing integration and expansion in $\varphi$ one gets
\be
\Omega^r_{\Delta\varphi} =\frac{\Delta^2}{2G^2_2}
-N(0)\Delta^2\left(\frac32-\ln\frac{\Delta}{\kappa}\right)
-N(0)\frac{\varphi^4}{4\mu^2}.\label{39}\ee We see that correction
of the order $\varphi^2$ is absent.

\section{Assembling the Pieces}

The main result of the paper is the Ginzburg-Landau functional
with both condensates included. It reads
$$\Omega(\Delta,\varphi;\mu, T)=\varphi^2 T^2_c\left
(\frac{T-T_c}{T_c}\right)+
+\varphi^4\frac{9}{2\pi^2}\left\{\int^{\Lambda/T_c}_0
dt\frac{tht}{t}\left(\frac{cht}{cht-ch\frac{\mu}{T_c}}\right)-1\right\}+
$$
\be
\Delta^2 N(0)\left(\frac{T-T_c}{T_c}\right) +\Delta^4
\frac{N(0)}{T^2_c} \frac{7\zeta(3)}{16\pi^2}.\label{40}\ee

The first two terms stem from the "chiral" part of the
thermodynamic potential while the "mixed" part resulted in the
last two terms with $\varphi$ contribution suppressed (only terms
of the order higher than $\varphi^2$ may appear). Comparing the
coefficients of the first two terms with those of the second two
one  notices that the contribution of the
 chiral condensate is damped by a factor $T_c^2/\mu^2\ll 1 (
 T_c\simeq 40$ MeV, $\mu\simeq  400$ MeV).

 Therefore we arrive to the  following two conclusions;

 (i) The system of gap equations for chiral and diquark
 condensates  actually decouples.

 (ii) As soon as diquark condensate is formed the contribution of
 the chiral condensate to thermodynamic quantities becomes
 strongly suppresses.

 In  analogy with the BCS theory these statements form  what may
 be somewhat loosely called the Anderson theorem.

 Its validity is supported not only by numerical calculations
 within several modes \cite{3}-\cite{5} but also within the random
 matrix approach \cite{7a} based solely on the symmetry of the
 system.

 Several assumptions and   simplifications made on the way to the
 above conclusions were outlined in the text. They are the
 following. The current quark masses were neglected and the same
 critical temperature was taken for chiral and diquark
 condensates. The first assumption is a technical one and was made
 in order to avoid lengthy equations. It will be lifted in the
 next publication. The equality of the critical temperatures
 $T_\varphi$ and $T_\Delta$ in the vicinity of the $\mu$ value
 where the diquark condensate arises was demonstrated by numerical
 calculations in \cite{3}. This  assumption is neither crucial and
 our final result (\ref{40}) is easily generalized to the case of
 different $T_\varphi$ and $T_\Delta$. The expression (\ref{9})
 for the thermodynamic potential is quite general and is valid for
 a variety of the interaction models and different values of the
 coupling constants. The central step is the expansion of the
 logarithm in (\ref{9}) up to $\Delta^2$ and $\Delta^4$ terms.
 This procedure is valid near the transition point but beyond the
 region of strong fluctuations.

 The suppression of the chiral condensate by the factor $T^2_c/\mu^2$
 in the final expression (\ref{40}) for $\Omega$ is due to the $\mu^2$ factor  in the
 density of quark states at finite chemical potential (see
 (\ref{27})).

  Finally we want to
 stress the point which we consider most important for future
 amendment. This is the investigation of the role played by gluon
 fields \cite{11}. Nonperturbative gluon fields introduce into the
 subject a new quantity - the correlation length $T_g$. It might be
 even speculated that the term $\Delta^3$ in the Ginzburg-Landau
 functional will appear.

 The author is grateful to Yu. A.Simonov, V.I.Shevchenko  and N.O.Agasian
 for discussions. Support from grants RFFI-00-02-17836, RFFI-
 00-15-96786 and INTAS CALL 2000 project \# 110 is gratefully acknowledged

\end{document}